\documentclass[a4paper]{article}

\usepackage{4d-validation-data-descriptor}

\title{An in vivo validation dataset for dynamic volumetric MRI}

\date{}

\begin{document}

\author[1]{Max H.C. van Riel\thanks{Corresponding author: m.h.c.vanriel-3@umcutrecht.nl}}
\author[1]{David G.J. Heesterbeek}
\author[2]{Martijn Froeling}
\author[1]{Cornelis A.T. van den Berg}
\author[1]{Alessandro Sbrizzi}
\affil[1]{Computational Imaging Group for MRI Therapy \& Diagnostics, Department of Radiotherapy, University Medical Center Utrecht, Utrecht, The Netherlands}
\affil[2]{Precision Imaging Group, Department of Radiology, University Medical Center Utrecht, Utrecht, The Netherlands}

\maketitle

\begin{abstract}
Dynamic volumetric MRI provides valuable information on in vivo motion and biomechanics, with applications spanning cardiac, musculoskeletal, or pulmonary imaging, amongst others.
Developing reconstruction methods for time-resolved volumetric MRI is challenging due to the inherently slow acquisition process of MRI, which makes it an active area of research.
However, in vivo validation of these methods remains challenging due to the lack of publicly available datasets with fully sampled ground-truth images.
Here, we present a publicly available in vivo dataset designed to facilitate the development and validation of dynamic volumetric MRI reconstruction algorithms.
Controlled and repeatable deformations of the muscles in the thigh were induced using a pneumatic pressure cuff, enabling the acquisition of both undersampled dynamic data and fully sampled validation images.
The dataset comprises multichannel undersampled k-space data from nine healthy volunteers across four different dynamic deformations, with fully sampled validation data for one deformation.
Additionally, an anatomical reference scan and muscle segmentation masks are provided for each subject.
To illustrate a possible image reconstruction and validation approach, a binning-based reconstruction was performed on the undersampled data from six dynamic repetitions.
The resulting images were consistent with the corresponding fully sampled validation images.
This dataset offers possibilities for validating and advancing time-resolved volumetric MRI reconstruction methods.

\end{abstract}

\section*{Background \& Summary}
Magnetic resonance imaging (MRI) is a widely-used imaging modality with excellent soft-tissue contrast. 
It enables noninvasive assessment of anatomical and physiological properties of the human body.
Conventional MRI acquisitions are typically performed under static conditions, where any motion is considered an artifact that has to be mitigated \cite{zaitsev_motion_2015}.
However, dynamic imaging and motion quantification can also serve as a valuable source of information.
Dynamic MRI techniques aim to capture and quantify in vivo motion and biomechanics, with a broad range of clinical and research applications. 
For instance, dynamic MRI of the heart enables the estimation of biomarkers such as left ventricular ejection fraction and myocardial strain, which are critical for diagnosing and monitoring cardiovascular diseases \cite{puntmann_society_2018, amzulescu_myocardial_2019}.
Similarly, dynamic MRI can be used to characterize muscle tissue properties \cite{sinha_age-related_2015, mazzoli_accelerated_2018}, assess pulmonary ventilation and perfusion \cite{ilicak_dynamic_2023}, investigate motion of the vocal tract for speech MRI \cite{scott_speech_2014}, and capture gastrointestinal motility \cite{de_jonge_evaluation_2019, lu_automatic_2022}.

A key challenge in dynamic MRI is the inherently slow acquisition process of MRI.
MRI scanners measure data in the spatial frequency domain, called k-space.
The signal in k-space needs to be sampled at the Nyquist rate to prevent aliasing artifacts.
Only after all required k-space samples have been acquired can an image be reconstructed using the inverse Fourier transform.
Acceleration of the data acquisition by undersampling the data, combined with advanced reconstruction techniques, is crucial to reduce scan time for static imaging, or to increase the temporal resolution for dynamic imaging.
These acceleration strategies include parallel imaging using multiple receive coils, compressed sensing using sparsity-promoting priors such as total variation or wavelet transforms, and low-rank constraints \cite{pruessmann_sense_1999,griswold_generalized_2002,lustig_sparse_2007,feng_goldenangle_2014,otazo_low-rank_2015}.

A common problem for dynamic 2D (2D+t) methods is through-slice motion, since most physiological motion is inherently three-dimensional.
Therefore, dynamic volumetric (3D+t) imaging methods are preferred, since they are capable of completely capturing the dynamic processes.
However, acquiring fully sampled 3D+t data at sufficient spatio-temporal resolutions is not feasible.
Current approaches often acquire data over multiple motion cycles, combining data from these repetitions into several motion bins to reconstruct a single representative cycle \cite{tan_motion-compensated_2023,chen_abdominal_2025,yerly_high_2025}.
This binning strategy requires consistent periodic motion, but achieving accurate repeatability of the motion is complicated in many dynamic imaging applications.
In cardiac applications for example, this is problematic when arrhythmias occur, or when patients have difficulties holding their breath \cite{contijoch_future_2024}.
Accurate repeatability of the motion is also complicated for other dynamic imaging applications, such as dynamic muscle imaging, where accurate repetitive motion is hard to achieve.
Therefore, the development of robust acquisition and reconstruction techniques for time-resolved 3D+t MRI that do not rely on the periodicity of the motion remains an active area of research \cite{ong_extreme_2020,feng_4d_2023,olausson_free-running_2025}.

When developing 3D+t MRI methods, it is important to have access to ground-truth data to validate the reconstruction results.
Currently, there are some 2D+t datasets available that provide fully sampled data, which can be retrospectively undersampled \cite{chen_ocmr_2020,wang_cmrxrecon_2024}.
The reconstruction from the undersampled data can then be compared to the ground-truth reconstruction using the fully sampled data.
However, there is a lack of publicly available 3D+t datasets with fully sampled ground-truth images that enable the validation of 3D+t reconstruction methods.

In this work, we present a publicly available in vivo dataset designed to support research in 3D+t MRI reconstruction methods.
We have created a controllable setup to induce repeatable deformations in the thigh muscles.
The dataset contains multichannel undersampled k-space data acquired from nine healthy volunteers.
Four distinct dynamic deformations are included for each subject.
For the first of these deformations, fully sampled data is also available, enabling validation of the reconstructed 3D+t images.
In addition to dynamic image reconstruction, the dataset enables analysis of strain and stiffness, as demonstrated in prior work for a single subject \cite{van_riel_4d_2025,heesterbeek_quantitative_2025}.

\section*{Methods}
\begin{figure}[!t]
    \centering
    \includegraphics[width=\textwidth]{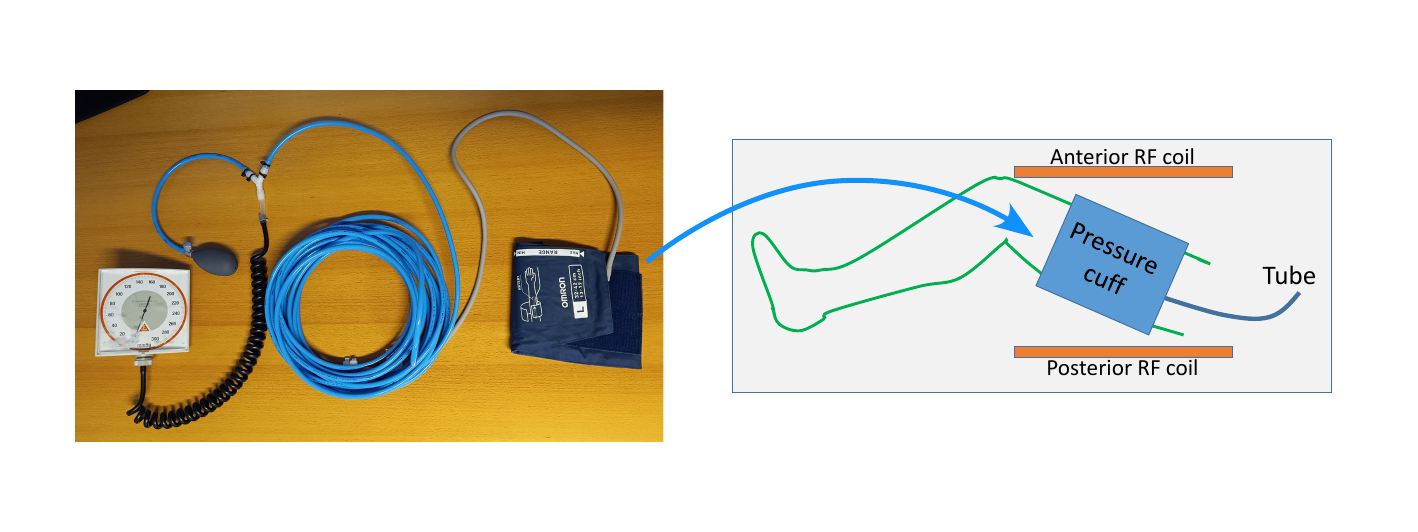}
    \caption{Experimental setup, from left to right: pressure gauge, hand pump, long tube (allowing operation outside of the scanner room), and pressure cuff, which is placed around the thigh of the volunteer.}
    \label{fig:exp_setup}
\end{figure}

\subsection*{Experimental setup}
To enable the acquisition of validation data, a setup that allows for controllable deformations was realized.
A pneumatic cuff from a manual blood pressure monitor was positioned around the right thigh of a healthy volunteer (\cref{fig:exp_setup}).
The pressure within the cuff was regulated using a manual pump and monitored via an analog pressure gauge (in mmHg).
Gradual inflation of the cuff applied an external compressive force on the subject's thigh, inducing time-dependent deformations of the muscles.
A long plastic pneumatic tube connected the cuff to the pump and the pressure gauge, allowing for remote operation outside the MRI scanner room.
This simple setup enabled the acquisition of undersampled data in a dynamic setting, as well as fully sampled validation images, as explained below.

\subsection*{MRI data acquisition}

\begin{table*}[!t]
\caption{Overview of the scans acquired for each subject.\label{tab:scans}}
\centering
\begin{tabular}{lllll}
    \toprule
    \textbf{Scan name} & \textbf{Pressure profile} & \textbf{Muscle condition} & \textbf{Cuff} & \textbf{Acquisition}\\
    & & & \textbf{position} & \textbf{time} (s) \\
    \midrule
    Anatomical reference & No pressure & Passive & N.A. & 123.7\\
    Dixon scan & No pressure & Passive & N.A. & 114.4\\
    Validation scan & 9 pressure levels & Passive & Initial & 199.1\\
    Hybrid scan  & 1 pressure cycle with & Passive & Initial & 66.4\\
    & sustained pressure & & \\
    Dynamic scan 1 & 6 pressure cycles & Passive & Initial & 221.2\\
    Dynamic scan 2 & 1 pressure cycle & Isometric knee flexion & Initial & 44.2\\
    Dynamic scan 3 & 1 pressure cycle & Passive & Rotated & 44.2\\
    Dynamic scan 4 & 1 pressure cycle & Isometric knee flexion & Rotated & 44.2\\
    \bottomrule
\end{tabular}
\end{table*}

All MRI imaging was performed on a 1.5T MRI system (Ingenia, Philips Healthcare, Best, The Netherlands).
The study was approved by the Institutional Review Board of the UMC Utrecht (NL53099.041.15, METC 15/466).
All data acquisition and analysis procedures were performed in accordance with the relevant regulations and guidelines of the UMC Utrecht.
Nine healthy volunteers (3F/6M) were included after giving informed consent for anonymized data sharing.
Subjects were placed in supine (face-up), feet-first position in the scanner, with the pressure cuff placed around their right thigh.
Their knees were supported to minimize bulk motion during and between the scans.
A 16-channel anterior coil and a 12-channel posterior coil (integrated in the table) were used for data acquisition.

For each subject, the following scans were acquired (\cref{tab:scans}): one anatomical reference scan, one Dixon scan, one validation scan, one hybrid dynamic and validation scan, and four dynamic scans.
The acquisition parameters are listed in \cref{tab:scan_param}.
All scans (apart from the Dixon scan) were acquired with an echo time (TE, i.e. the time between the radiofrequency pulse and the readout) of 2.30 ms.
At this particular echo time, water and fat have opposite phase, producing dark bands at muscle boundaries that enhance segmentation and motion detection.

\subsubsection*{Anatomical reference and segmentation}
As an anatomical reference, a high-resolution static scan was acquired without deformations.
This scan shows finer anatomical details than the dynamic data.
Additionally, a Dixon scan was acquired with three echo times (1.20 ms, 2.67 ms, and 4.14 ms), from which muscle segmentation masks could be generated.

\subsubsection*{Validation data}

\begin{table*}[!t]
\caption{Acquisition parameters for each scan.\label{tab:scan_param}}
\centering
\begin{tabular}{lllll}
    \toprule
    & \textbf{Anatomical} & \textbf{Dixon} & \textbf{Validation} & \textbf{Hybrid \&} \\
    & \textbf{reference} & \textbf{scan} & \textbf{scan} & \textbf{Dynamic scans}\\
    \midrule
    \textbf{Scan technique} & 3D RF-spoiled & 3D gradient echo & 3D RF-spoiled & 3D RF-spoiled \\
    & gradient echo & & gradient echo & gradient echo \\
    \textbf{Sampling} & Linear & Linear & Linear & CASPR \\
    \textbf{Repetition time} (ms) & 5.40 & 15.0 & 5.40 & 5.40 \\
    \textbf{Echo time} (ms) & 2.30 & 1.20/2.67/4.14 & 2.30 & 2.30 \\
    \textbf{Flip angle} (\degree) & 6 & 6 & 6 & 6\\
    \textbf{Resolution} (mm\(\times\)mm\(\times\)mm) & 1.4\(\times\)1.4\(\times\)1.5 & 1.5\(\times\)1.5\(\times\)6.0 & 3.5\(\times\)3.5\(\times\)3.5 & 3.5\(\times\)3.5\(\times\)3.5 \\
    \bottomrule
\end{tabular}
\end{table*}

\begin{figure}[!t]
    \centering
    \includegraphics[width=0.75\textwidth]{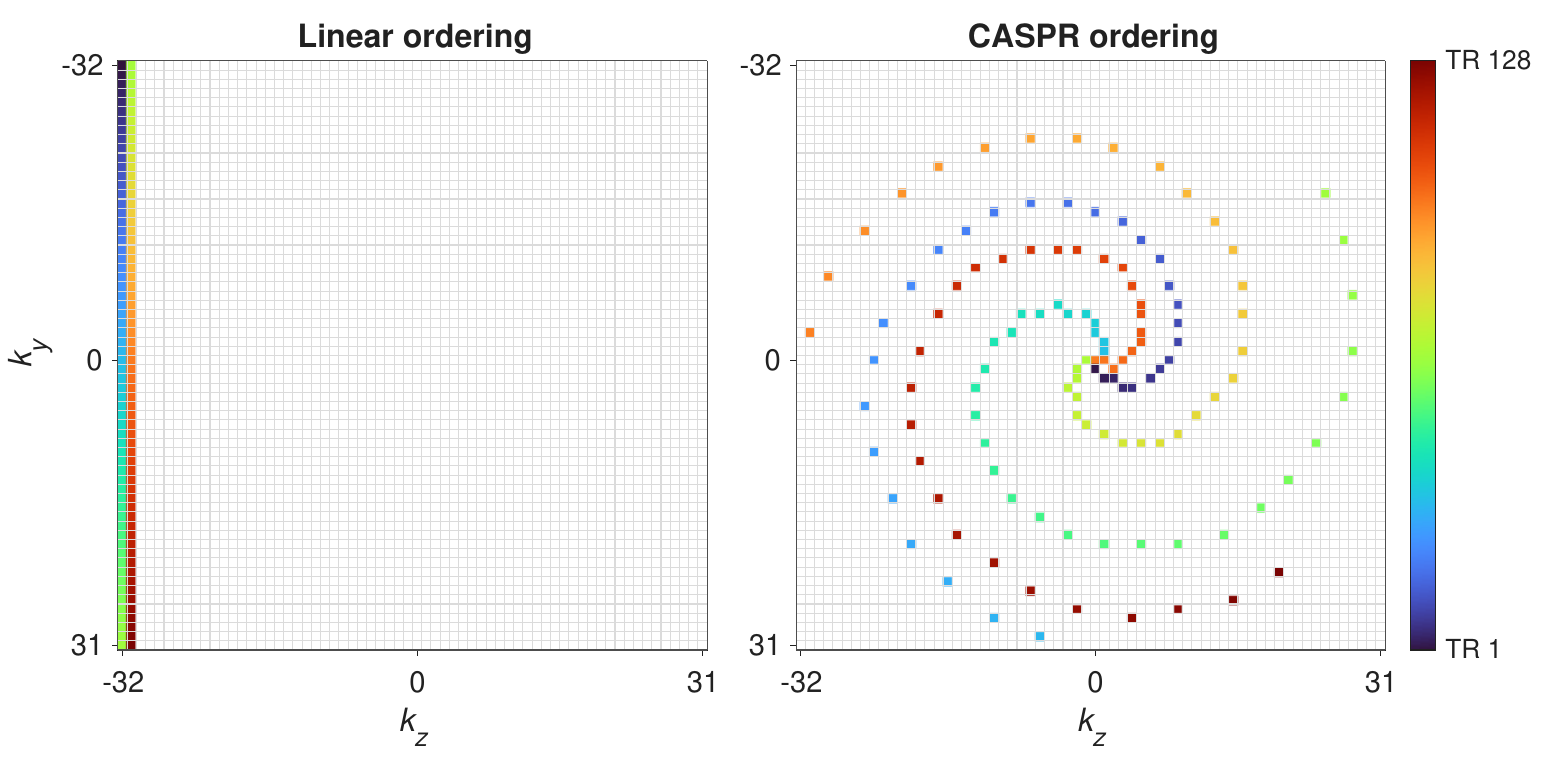}
    \caption{Left: Regular linear sampling pattern, as used for the static validation data. Right: CASPR sampling pattern, as used for the dynamic data and the hybrid dynamic/validation data. The readout direction (\(k_x\), not shown) is perpendicular to the figure. The colors indicate the sampling order for the first 128 readouts.}
    \label{fig:sampling}
\end{figure}

Fully sampled validation images were acquired at nine discrete pressure levels.
The pressure cuff was inflated from 0 to 80 mmHg and deflated back to 0 mmHg in 20 mmHg increments, where the pressure was kept constant at each increment for 22 seconds to acquire a fully sampled validation image (\cref{fig:pressures}).
The k-space was sampled on a grid of 128-by-64-by-64 with an isotropic spatial resolution of 3.5 mm.
During each TR, one readout line (128 samples) along the \(k_x\)-direction was acquired. The \(k_y\)-\(k_z\) plane was sampled using a conventional linear sampling pattern (\cref{fig:sampling}).

In addition, a single repetition of the pressure cycle was performed: the pressure cuff was continuously inflated from 0 mmHg (no deformation) to 80 mmHg (maximum deformation), where the pressure was sustained for 20 seconds, and subsequently deflated back to 0 mmHg.
This resulted in a hybrid dataset containing both dynamic and validation data, where the dynamic part can be used for a time-resolved reconstruction, whereas the data acquired while the pressure was sustained can be used to generate one validation image at maximal deformation.
The k-space was sampled on the same grid as the validation data, but during this scan, the readouts were sampled using Cartesian acquisition with spiral profile order (CASPR) \cite{prieto_highly_2015} in the \(k_y\)-\(k_z\) plane, with 32 readouts over a rotation of 360\(\degree\) per spiral shot (\cref{fig:sampling}).
This sampling strategy combines the benefits of non-Cartesian sampling, like the oversampling of the k-space center, with the robustness of Cartesian sampling against field inhomogeneities, gradient delays, and eddy currents.

\subsubsection*{Dynamic data}
During the dynamic scans, the pressure cuff was inflated from 0 mmHg (no deformation) to 80 mmHg (maximum deformation) and immediately deflated back to 0 mmHg.
This pressure cycle resulted in repeatable elastic deformations of the muscles.

\begin{figure}[!t]
    \centering
    \includegraphics[width=\textwidth]{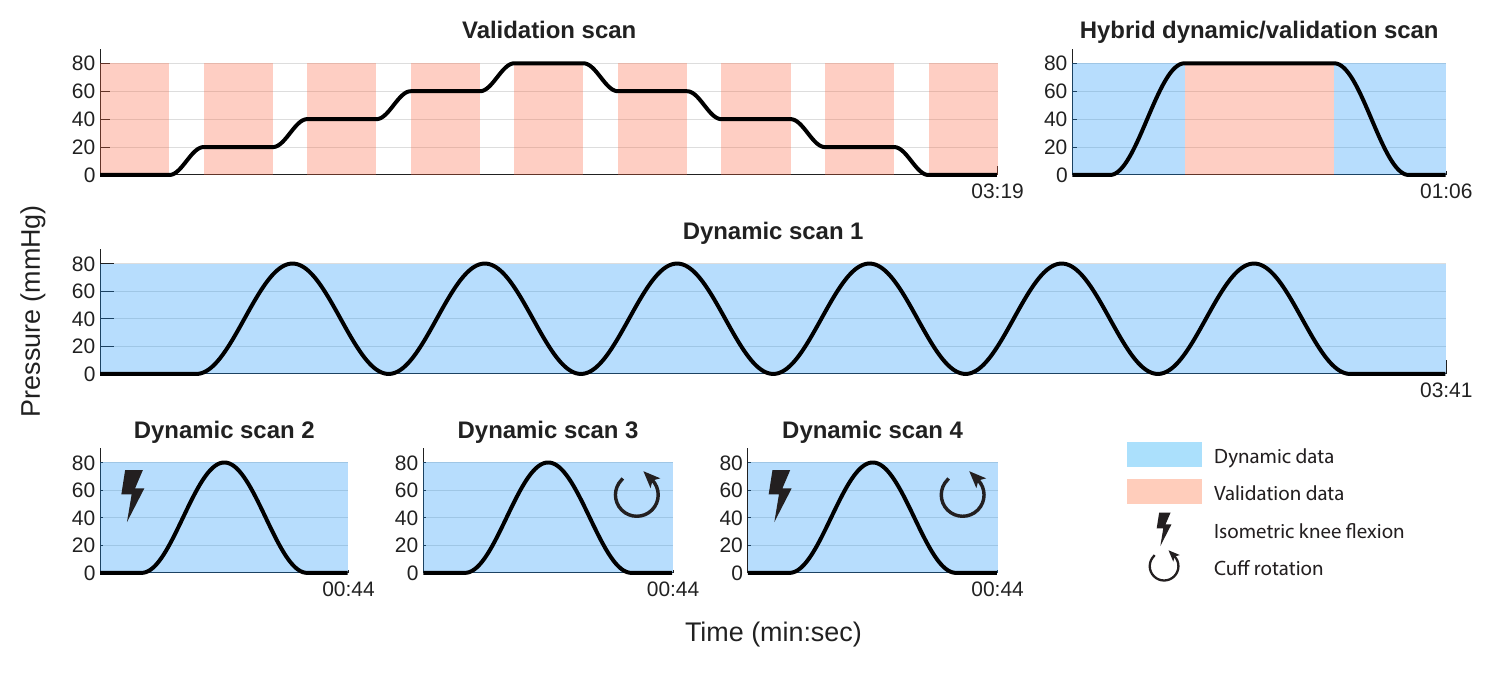}
    \caption{Pressure profiles for the validation scans and the dynamic scans. Data acquired during the blue-shaded time intervals can be used for dynamic time-resolved reconstructions, while data acquired during the orange-shaded time intervals can serve as validation data for the dynamic reconstructions. Dynamic scans 2 and 4 were acquired during an isometric knee flexion task. During dynamic scans 3 and 4 the pressure cuff was rotated. The acquisition time for each scan is indicated on the horizontal axis.}
    \label{fig:pressures}
\end{figure}

For each subject, four dynamic scans were performed (\cref{fig:pressures}):
\begin{enumerate}
    \item Six consecutive repetitions of the pressure cycle at 35-second intervals.
    \item A single pressure cycle, during which the volunteer was asked to press their heel into the scanner table. This activated the muscles in the posterior (back) side of the leg, which are responsible for bending the knee joint, leading to isometric flexion of the knee.
    \item The pressure cuff was repositioned at a different angle around the thigh and a single pressure cycle was performed. This changed the pressure distribution around the leg, thus changing the deformation patterns in the muscles.
    \item The isometric knee flexion task of dynamic scan 3 was repeated with the rotated pressure cuff.
\end{enumerate}

All four dynamic scans were acquired using the  CASPR sampling pattern (\cref{fig:sampling}).

\subsection*{Image reconstruction}
For the scans with fully sampled data, an image was reconstructed using the inverse Fourier transform followed by a sensitivity-weighted combination over the receive channels:

\begin{equation}
\rho(\mathbf{x}) = \frac{1}{\sum_l^{N_c} c_l^*(\mathbf{x}) c_l(\mathbf{x})} \sum_j^{N_c} c_j^*(\mathbf{x}) \mathcal{F}^{-1} [s_j(\mathbf{k})],
\label{eq:direct_recon}
\end{equation}

\noindent with \(\rho(\mathbf{x})\) the image intensity at location \(\mathbf{x}\), \(s_j(\mathbf{k})\)the k-space signal from the \(j\)-th channel at k-space coordinate \(\mathbf{k}\), \(\mathcal{F}^{-1}\) the inverse Fourier transform, \(c_j(\mathbf{x})\) the coil sensitivity of the \(j\)-th receive channel, \(N_c\) the number of receive channels, and \(^*\) indicating complex conjugation.
The coil sensitivity maps were determined from a separate prescan and are also included in the dataset.

\subsection*{Muscle segmentations}
The out-of-phase images from the Dixon scan were converted from Digital Imaging and Communications in Medicine (DICOM) to Neuroimaging Informatics Technology Initiative (NIfTI) file format using dcm2niix v1.0.20250505 \cite{li_first_2016}.
Muscle segmentations were generated from these NIfTI files using a pretrained U-net \cite{rohm_3d_2021} as implemented in QMRITools 4.2.8 \cite{froeling_qmrtools_2019}. These segmentation masks were visually checked and where necessary corrected using ITK-SNAP 4.0.1.

\section*{Data Records}
The data \cite{van_riel_vivo_2026} can be found at \href{https://www.doi.org/10.5281/zenodo.21194933}{doi.org/10.5281/zenodo.21194933}.
All datasets are saved in the MRD file format \cite{inati_ismrm_2017}.
For each subject, the following files are available:
\begin{itemize}
    \item anatomy.nii.gz: High-resolution fully sampled anatomical reference image
    \item outphase.nii.gz: Out-of-phase image from the Dixon scan
    \item segmentation.nii.gz: Muscle segmentation masks as determined from the out-of-phase image; each voxel is assigned an integer corresponding to a certain segmentation label
    \item segmentation.txt: Label descriptions of the segmentation masks
    \item validation.mrd: Fully sampled multichannel k-space data for each of the nine pressure levels
    \item hybrid.mrd: Undersampled multichannel k-space data for the hybrid dynamic and validation scan
    \item dynamic1.mrd, \ldots, dynamic4.mrd: Undersampled multichannel k-space data for the four dynamic scans
\end{itemize}
The only preprocessing performed to the raw k-space data is the removal of the oversampling in the readout direction, as this reduces the size of the dataset by a factor 2.
In addition to the k-space data, each data file containing raw k-space data also contains noise measurements, required for prewhitening the data.
Each .mrd file is a HDF5 file that contains the raw k-space and noise data in MRD format in \texttt{/dataset/data}, the MRD header in \texttt{/dataset/xml}, and the coil sensitivity maps in \texttt{/csm}.
The encoding counters stored in the MRD AcquisitionHeader contain the \(k_y\) and \(k_z\) coordinates of each readout line, specifying the exact sampling pattern for each scan.
This information is especially important for the reconstruction of the CASPR data.
The nine pressure levels in the validation dataset are indexed using the repetition counter in the MRD AcquisitionHeader.
The first user-defined counter in the MRD AcquisitionHeader is used to distinguish the six repetitions in dynamic scan 1, and to indicate the static part of the hybrid scan.

\section*{Data Overview}
\begin{figure}[!t]
    \centering
    \includegraphics[width=0.75\textwidth]{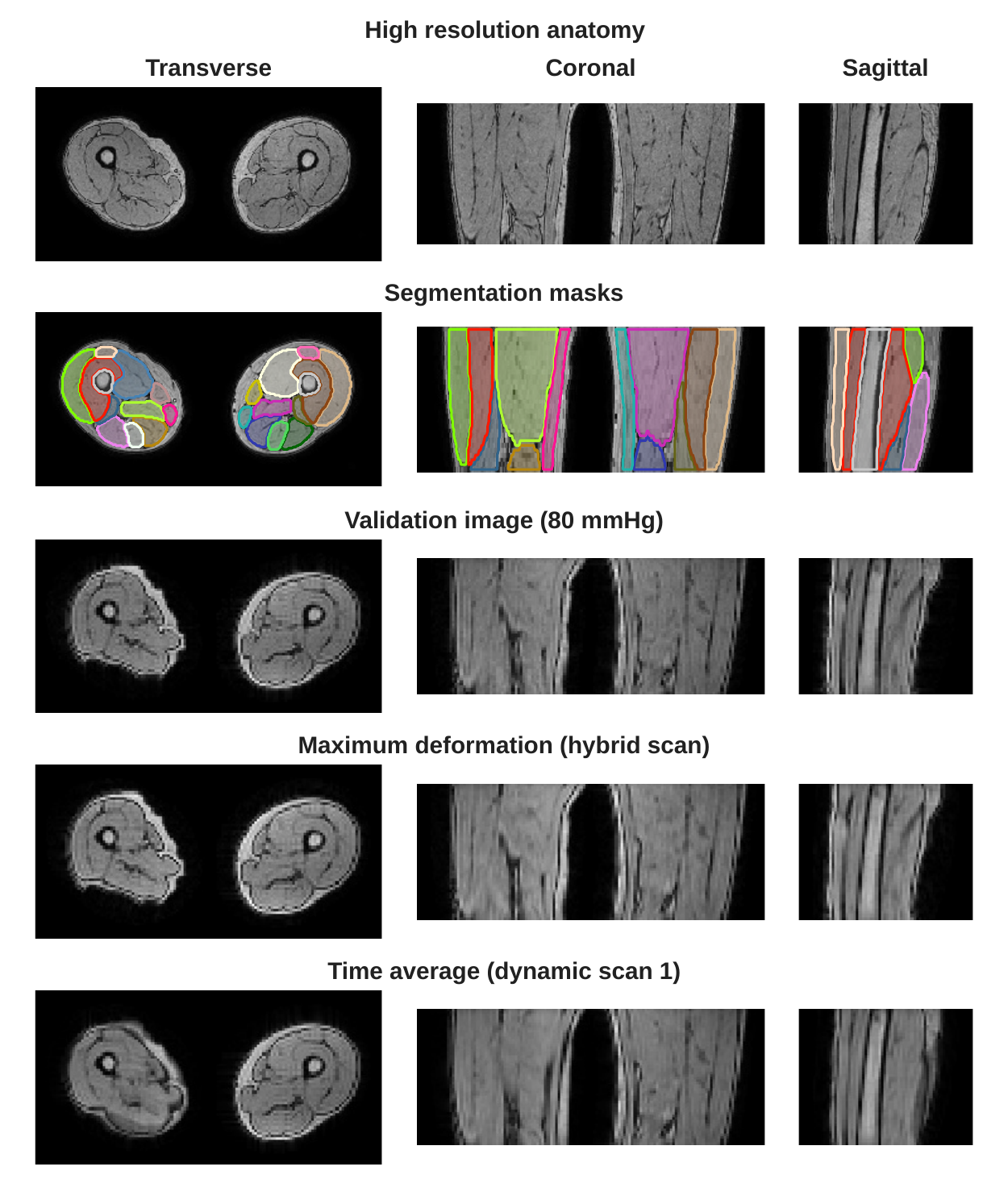}
    \caption{Overview of the different scans in the dataset. From top to bottom: high-resolution anatomical image; out-of-phase image from the Dixon scan with the muscle segmentation masks overlaid; fifth validation image from the static validation data (at the maximum pressure level of 80 mmHg); image at the maximum deformation reconstructed from the validation part of the hybrid dynamic/validation scan; time-averaged image from the data of all six dynamic repetitions of dynamic scan 1. Note that due to the radiological convention, the right leg is shown on the left side of the transverse and coronal images.}
    \label{fig:data_overview}
\end{figure}

\Cref{fig:data_overview} shows representative images for a single volunteer that can be reconstructed from the dataset using the direct image reconstruction given by equation \labelcref{eq:direct_recon}.
Note that we used the radiological convention to display the images, which means that the right leg is shown on the left side of the image.

\section*{Technical Validation}
Direct reconstructions of time-resolved images from the undersampled k-space data would result in severe artifacts (like the motion artifacts in the time-averaged reconstruction in \cref{fig:data_overview}), unless advanced reconstruction algorithms are used, which is beyond the scope of this data descriptor.
Nevertheless, we can still reconstruct motion-resolved images from dynamic scan 1 using a binning strategy.
Note that this approach is not time-resolved, since it uses the data from multiple dynamic repetitions and thus assumes periodicity of the motion.

\begin{figure}[!t]
    \centering
    \includegraphics[width=\textwidth]{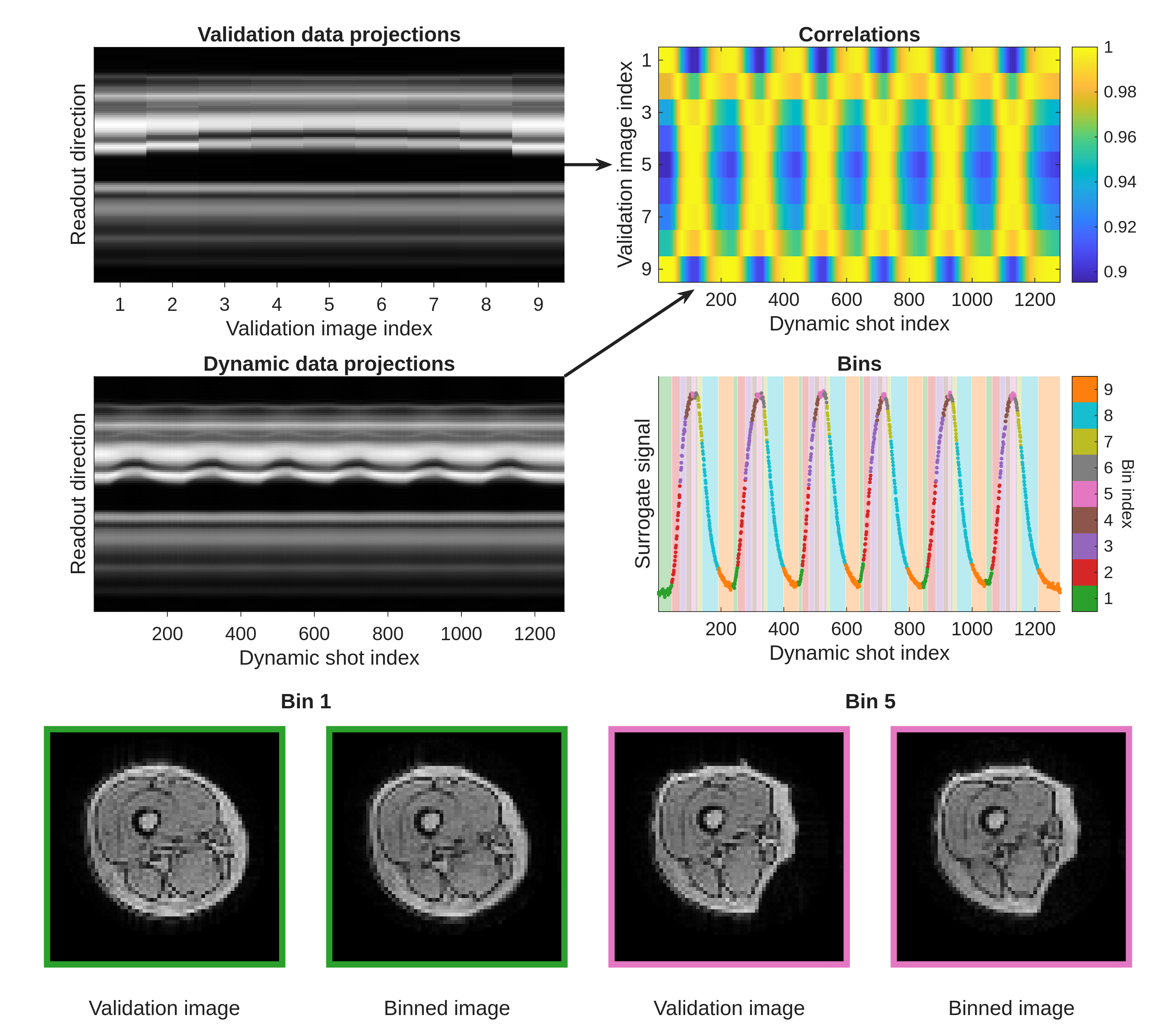}
    \caption{Binned image reconstruction approach. Projections on the readout axis are calculated for each validation image and each CASPR shot of dynamic scan 1. The correlations between these projections are then used to divide the shots into nine bins, with each bin corresponding to one validation image. The first principal component of the projection data is used as motion surrogate signal. The validation images and corresponding images from the binning strategy are shown for bin 1 (no deformation) and bin 5 (maximal deformation).}
    \label{fig:validation}
\end{figure}

Each pseudo-spiral shot of the CASPR sampling pattern begins by sampling the line through the center of k-space.
A 1D inverse Fourier transform yields the projection of the imaging data on the \(x\)-axis (the readout direction).
This projection changes as the image deforms, and thus can be used as a surrogate signal that shows the motion that happened during the scan.
Using the correlation between the projections of the dynamic scan and those of the validation scan, the data from the six repetitions of dynamic scan 1 can be grouped into nine bins, with each bin corresponding to one of the nine validation images.
Each bin is only slightly undersampled, allowing image reconstruction via a regularized parallel imaging reconstruction, as implemented in BART 0.9 \cite{blumenthal_mrireconbart_2023}.
The resulting images of this binned reconstruction show high similarity to the corresponding validation images (\cref{fig:validation}), confirming that the induced deformations during the dynamic scans were consistent with those observed in the validation scan.

\section*{Usage Notes}
For dynamic scan 1, periodicity of the deformation pattern across the six repetitions can be exploited during the reconstruction, as demonstrated in the Technical Validation section (\cref{fig:validation}).
Alternatively, this assumption can be discarded, and (a subset of) the data can be reconstructed in a time-resolved manner.
In particular, six independent reconstructions (one for each pressure cycle) can be performed to investigate the repeatability of the reconstruction method.
Optionally, the undersampling factor can be retrospectively increased by discarding selected spiral shots.
Reconstructions obtained from dynamic scan 1, whether binned or time-resolved, can be quantitatively compared with the validation data.

When combining data from different scans, users of the dataset should be aware that there may be minor positional shifts of the leg between scans.
This could be caused by the repositioning of the pressure cuff, or by the subject moving their leg during the isometric knee flexion tasks.
Therefore, coil sensitivities were determined for each scan separately, as these may be different across different scans.
Furthermore, when using the segmentation masks, a nonrigid registration between the out-of-phase Dixon image (on which the segmentation was performed) and an image of the target reconstruction (for example, the first frame) is recommended.
In addition, a minor misalignment might appear when comparing the validation images to reconstructed images from one of the dynamic scans.
Only rigid registration should be performed in this case to prevent warping the validation or reconstructed images.
Since both legs can move independently, we advise cropping out the right leg in both images, followed by a rigid registration.

The left leg does not deform, and can thus be used as a control to compare against.
Note the absence of motion artifacts in the left leg (right side of the image) in the time-averaged image in \cref{fig:data_overview}, and the constant projections in \cref{fig:validation}.

The hybrid dynamic and validation data can be split into two subsets: a dynamic part for a time-resolved reconstruction, and a static part from which a single validation image can be reconstructed.
The advantage of this validation approach is that both subsets correspond to the same pressure cycle, eliminating the need for registration between scans.
However, validation using this strategy is limited to the deformation state at the maximum pressure level.

For the remaining three dynamic scans, no validation data is available.
These scans are intended to investigate different deformation patterns and the effect of varying biomechanical conditions on the tissue motion.

\section*{Data availability}
The dataset is publicly available on Zenodo at \href{https://www.doi.org/10.5281/zenodo.21194933}{doi.org/10.5281/zenodo.21194933}.

\section*{Code availability}
An example script that loads the data, preprocesses the raw data, and performs a basic image reconstruction is included in the dataset.

\section*{Author contributions}
M.R., D.H., and A.S. developed the methodology and acquired the data. M.R. and M.F. analyzed the data. A.S. and C.B. supervised the project. M.R. wrote the main manuscript text. A.S. acquired the funding for this project. All authors reviewed the manuscript.

\section*{Funding}
This research was supported by the Dutch Research Council (NWO), grant 18897.

\end{document}